\documentclass{elsart}
\usepackage{graphicx}

\newcommand{\req}[1]{(\ref{#1})}
\newcommand{\be}{\begin{equation}}
\newcommand{\ee}{\end{equation}}
\newcommand{\bea}{\begin{eqnarray}}
\newcommand{\eea}{\end{eqnarray}}

\newcommand{\avg}[1]{\langle{#1}\rangle}

\newcommand{\BE}{\begin{eqnarray}}
\newcommand{\EE}{\end{eqnarray}}
\newcommand{\BEn}{\begin{eqnarray*}}
\newcommand{\EEn}{\end{eqnarray*}}
\newcommand{\barr}{\begin{array}}
\newcommand{\earr}{\end{array}}

\newcommand{\bit}{\begin{itemize}}      
\newcommand{\eit}{\end{itemize}}
\newcommand{\bc}{\begin{center}}
\newcommand{\ec}{\end{center}}
\newcommand{\ben}{\begin{enumerate}}    
\newcommand{\een}{\end{enumerate}}

\newcommand{\eps}{\epsilon}

\begin{document}

\begin{frontmatter}
\title{Stylized facts in minority games with memory: a new challenge}
\author{Damien Challet} \address{Nomura Centre for Quantitative
Finance, Mathematical Institute, Oxford University, 24--29 St Giles',
Oxford OX1 3LB, United Kingdom} \author{Matteo Marsili} \address{The
Abdus Salam International Centre for Theoretical Physics, Strada
Costiera 11, 34100 Trieste, Italy} \author{Andrea De Martino}
\address{INFM-SMC and Dipartimento di Fisica, Universit\`a di Roma
``La Sapienza'', P.le A. Moro 2, 00185 Roma, Italy}
\begin{abstract}
A finite memory is introduced in the score dynamics of Minority
Games. As expected, this removes the dependence of the stationary
state on the initial conditions. However, it also causes an unexpected
increase of fluctuations in grand-canonical models for very large
times. Current analytical methods are inadequate to solve this simple
and natural extension.
\end{abstract}
\begin{keyword} Minority Game \sep financial markets \sep finite
  memory \sep volatility clustering \sep fat tails \sep price return

\PACS 
\end{keyword}
\end{frontmatter}


\section{Introduction}

Remarkable new phenomenology arises when agents are granted the
possibility of not participating in a game \cite{Hauert,CM03}. In the
context of the Minority Game (MG) \cite{CZ97}, this extension -- which
takes the name of Grand Canonical MG (GCMG) -- has been introduced
independently by several authors \cite{SZ99,J99,MMM}. The MG captures
some aspects of financial market dynamics, namely the interplay
between exploitable predictability, price fluctuation and trader
behaviour (we refer the interested reader to
\cite{CM03,BouchMG2,Sornette,Book} for more details). Furthermore,
GCMGs reproduce some of the main stylized facts of financial market
phenomenology, such as large price changes and volatility clustering
\cite{CM03,J99,J00,CMZ00}. Remarkably, these features emerge close to
a phase transition, thus suggesting that financial markets operate in
the vicinity of a critical point \cite{CM03,Book}.

For physicists, this is almost intuitive. Nevertheless, it is hard to
convince economists that it is actually true. The MG is able to
propose a coherent account of why financial markets are at
criticality, based on market (in)efficiency (or predictability)
\cite{CM03,Book}. The idea of explaining anomalous fluctuations as a
critical phenomenon is not new. Other models, notably those based on
two-dimensional percolation \cite{ContBouchaud}, also relate stylized
facts with criticality. There is however little reason to consider the
market as bi-dimensional (although the trading floors are), or to
believe it should spontaneously reach the critical percolation
probability. The fact that all kinds of financial markets generate
power-law-distributed price changes implies that a reasonable model of
a financial market should not need the fine-tuning of a parameter.

An annoying feature of the GCMG is that, close to the phase
transition, its dynamics displays a peculiar dependence on initial
conditions: anomalous fluctuations only materialize in some
realizations of the dynamics, whereas others exhibit plain Gaussian
price fluctuations \cite{CM03}. The introduction of a finite memory in
the learning dynamics of agents, such that the impact that events
occurred far in the past have on the agents' choices in the present
vanishes with the time lag, is the natural remedy to this
inconvenient. This modification was introduced in \cite{MMRZ} for a
model in which agents play the MG strategically. Here we focus on the
standard MG, where agents adopt a na\"\i ve price-taking behavior
\cite{Book}.

We shall first show that indeed a finite memory eliminates the
dependence on initial conditions. A direct extrapolation of
theoretical results \cite{MMRZ} suggests that a finite memory leads to
an increase of fluctuations. This is confirmed by numerical
simulations. Then we shall see that, however, memory brings about new
non-trivial effects in the GCMG. In particular, it causes a build-up
of fluctuations in the stationary state that sets in after very large
times. Our discussion will be mainly informal and we shall give only
the minimal technical details. We will refer to original papers for
the detailed definition of the models.

\section{Finite memory score in the standard MG}

In few words, the MG models a system of $N$ interacting agents who
should repeatedly take a binary decision, such as buy/sell. Each agent
aims at taking the minority decision. This captures the fact that it
is generally convenient to buy when the majority sells and
vice-versa. Agents resort to trading strategies to process market
information. The latter takes one of $P=\alpha N$ possible values. The
performance of each strategy is monitored by a score function. The
dynamics of the score of strategy $s$ has the form \be
U_{i,s}(t+1)=U_{i,s}(t)-a_{i,s}^{\mu(t)}A(t)
\label{Uis}
\ee where $a_{i,s}^{\mu(t)}=\pm 1$ is the action prescribed by
strategy $s$ when the information takes the value $\mu(t)$ and $A(t)$
is the sum of all actions ($\pm 1$) of agents. At each time step,
agents adopt the strategy with the highest score. The last term in
(\ref{Uis}) is the payoff of agent $i$ for using strategy $s$ at time
$t$. As the particular form of the dynamics is not at stake in what
follows, we refer to \cite{Book} for more details. We just remind that
the key observables are the magnitude of fluctuations
$\sigma^2=\avg{A^2}$ and the information content $H=(1/P)\sum_\mu
\avg{A|\mu}^2$, which details how the outcome $A(t)$ is correlated
with the information $\mu(t)$ \cite{Book,CMZe99,MC01}. We also remind
that the value $N= P/\alpha_c$, with $\alpha_c=0.3374\ldots$,
separates an asymmetric, information rich phase with $H>0$ (for
$N<P/\alpha_c$) from a symmetric phase with $H=0$ ($N>P/\alpha_c$). In
the latter, the stationary state depends on the asymmetry
\[
U_0=U_{i,1}(0)-U_{i,2}(0)
\] 
of the initial conditions (we focus on the MG with two strategies per
agent). This dependence is due to the fact that, under \req{Uis}, all
market fluctuations since time $t=0$ are remembered and contribute
with the same weight to $U_{i,s}(t)$, irrespective of how far in the
past they took place.

This suggests to generalize (\ref{Uis}) to

\be
\label{scoreMGlambda}
U_{i,s}(t+1)=(1-\lambda/P)U_{i,s}(t)-a_{i,s}^{\mu(t)}A(t).
\ee

\begin{figure}
\centerline{\includegraphics[width=0.6\textwidth]{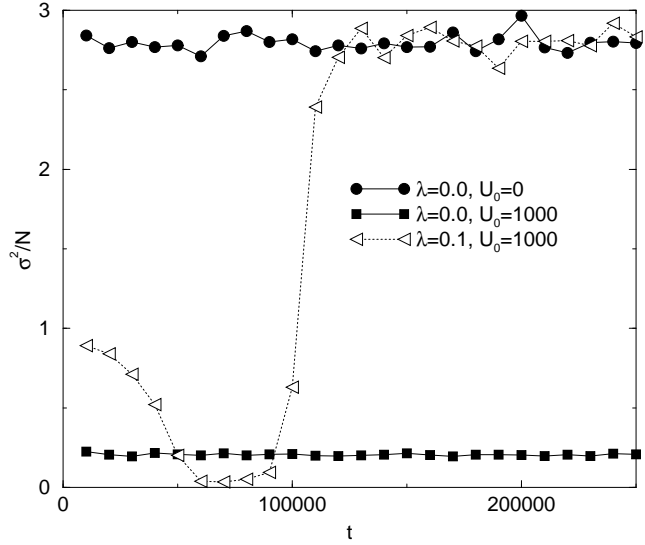}}
\caption{The same realization of the standard minority game with
  unbiased initial strategy scores and $\lambda=0$ (circles), biased
  initial scores and $\lambda=0$ (squares), biased initial scores and
  $\lambda=0.1$ (triangles).}
\label{s2t}
\end{figure}
\begin{figure}
\centerline{\includegraphics[width=0.6\textwidth]{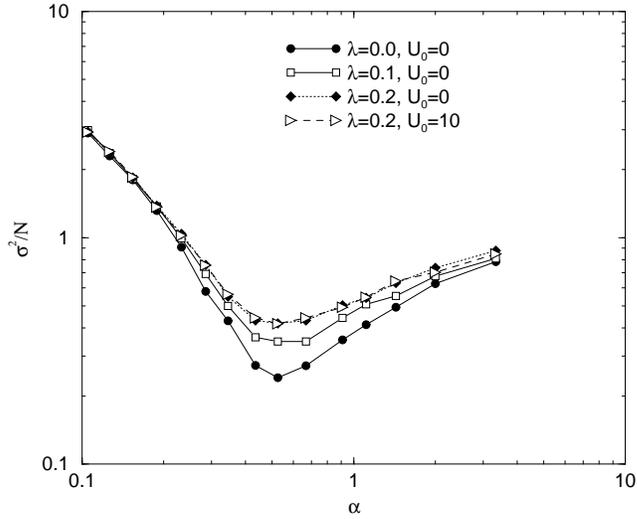}}
\caption{Average volatility $\sigma^2/N=\avg{A^2}/N$ versus
  $\alpha=P/N$ of the standard MG with unbiased initial scores and
  $\lambda=0$ (circles), $\lambda=0.1$ (squares), $\lambda=0.2$
  (diamonds), with biased initial scores and $\lambda=0.2$
  (triangles).}
\label{s2alpha}
\end{figure}
\begin{figure}
\centerline{\includegraphics[width=0.6\textwidth]{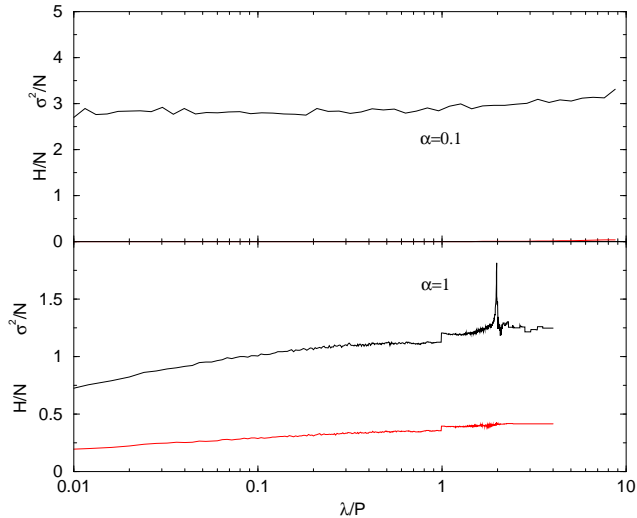}}
\caption{$\sigma^2/N$ (black lines) and $H/N$ (red lines) for a realization of the standard Minority Game with increasing $\lambda$}
\label{s2Hlambda}
\end{figure}

with $\lambda>0$ a constrant. Fig. \ref{s2t} shows that a time
dependent $\sigma^2=\avg{A^2}_t$~\footnote{Here $\avg{\ldots}_t$
stands for an average over a long but finite time interval around
$t$.} converges, for long times, to a value which is independent of
initial conditions $U_0$. This implies that the initial asymmetry
$U_0$ has no influence on the stationary state. This is confirmed by
Fig. \ref{s2alpha}, where we plot $\sigma^2/N$ as a function of
$\alpha=1/n_s$ for different $U_0$ and $\lambda$. For a fixed $\alpha$
and for a given realization of the game, $\sigma^2$ and $H$ are
increasing functions of $\lambda$, although in the symmetric phase
($\alpha<\alpha_c=0.3374\dots$) $\sigma^2$ varies very slowly with
$\lambda$ (see Fig. \ref{s2Hlambda}). While only small values of
$\lambda$ make physical sense, for completeness the same figure
reports also the queer effects of very large $\lambda$. From
\req{scoreMGlambda}, $\lambda=P$ cancels the contribution of $U_i(t)$
in $U_i(t+1)$; larger $\lambda$ implies that this contribution is of
opposite sign to the usual MG.

\section{Finite memory in the Grand Canonical MG}

\begin{figure}
\centerline{\includegraphics[width=0.6\textwidth]{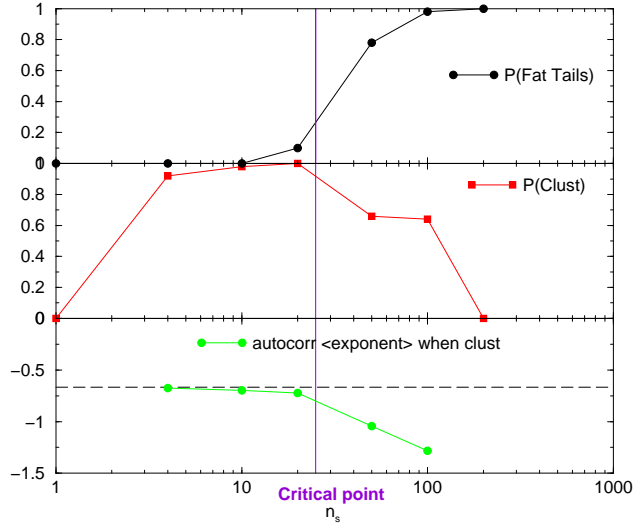}}
\caption{From top to bottom: fraction of the realizations of a grand
  canonical MG that has power-law tailed distribution of $A$,
  power-law autocorrelation function $\avg{|A(t)||A(t+\tau)|}
  \propto\tau^{-\gamma}$ (volatility clustering), and exponent
  $\gamma$ when volatility clustering is present ($P N_s =16000$,
  $n_p=1$, $\eps=0.01$, $10^7$ iterations, average over 50
  realizations).}
\label{volclust}
\end{figure}

Let us focus now on the simplest GCMG \cite{CM03}, where the agents ({\it
speculators}) possess a single strategy each and may abstain
from playing if the game is not profitable enough. A measure of
profitability is given by a parameter $\epsilon$, which measures the
gains from investing outside of the market. Again, we refer to the
original paper \cite{CM03} for a detailed account. The important new
feature of this model is that the number of participants is not fixed
a priori, but is dynamically determined by market gains. The model,
which includes a number $N_p$ of {\em producers} who trade no matter
what, according to a given strategy, can be solved exactly in the
limit of infinitely many speculators $N_s\to\infty$ (with
$\alpha=P/N_s=1/n_s$ and $n_p=N_p/P$ finite) via a static replica
calculation (this solution can be also recovered with the dynamical
generating functional technique \cite{CMM04}). This allows for a
systematic understanding of the model's properties. It turns out that
a phase transition occurs for $\epsilon=0$ at $n_s=1/\alpha> n_s^c$. The
transition is discontinuous across the $\epsilon=0$ line, where the
number of active agents jumps abruptly. Stylized facts are observed,
for $\epsilon>0$, close to the point $\eps=0$, $n_s\approx n_s^*(P)>n_s^c$ in
finite systems,\footnote{$n_s^*(P)$ increases with $P$ \cite{CM03}.} and the size of the critical region can also be
characterized. More precisely, we observe the occurrence of power-law
distribution of returns $A(t)$ and volatility clustering
\cite{BouchMG2}, or long-time correlations\footnote{Note that other
functions for describing the long-range correlation have been proposed
\cite{BouchMG2,Muzy}}
$\avg{|A(t)||A(t+\tau)|}\propto\tau^{-\gamma}$. These findings are
summarized in Fig. \ref{volclust}.
First we see that fat tails or volatility clustering both emerge close
to the phase transition $n_s\approx n_s^*(P)$ and $\epsilon\approx 0$. Then,
these two features do not appear in all runs. Some runs, even in the
critical region, show plain Gaussian returns, sometimes with long
range correlations and sometimes without. In the real world, this
would imply that some markets show fat-tailed price changes, and some
others not, which is at odds with the observed universality.

In the GCMG, this issue signals a non-ergodic behavior and is related
to the dependence on initial conditions. Both effects arise because of
the presence of infinite memory in the strategy scores dynamics. This
suggests to introduce a finite memory $\lambda>0$ in the
dynamics\footnote{A second solution may be to allow for the evolution
of agent's strategies. In that case, as observed in \cite{MMRZ}, an
infinite memory would make no sense.}, as done in the previous section
for the standard MG (see (\ref{Uis}) and (\ref{scoreMGlambda})). We
remark that, technically, finite score memory can also be implemented
by fixing a time window $T$ such that only the $T$ most recent scores
are kept by the agents, as in \cite{J99,CZ98}\footnote{Such games with
binary payoffs \cite{CZ97} can be written as a matrix; predicting
large price changes amounts to studying its eigenvalues
\cite{Jpred}.}. The two procedures are equivalent for $T\sim
P/\lambda$. 
\begin{figure}
\centerline{\includegraphics[width=0.6\textwidth]{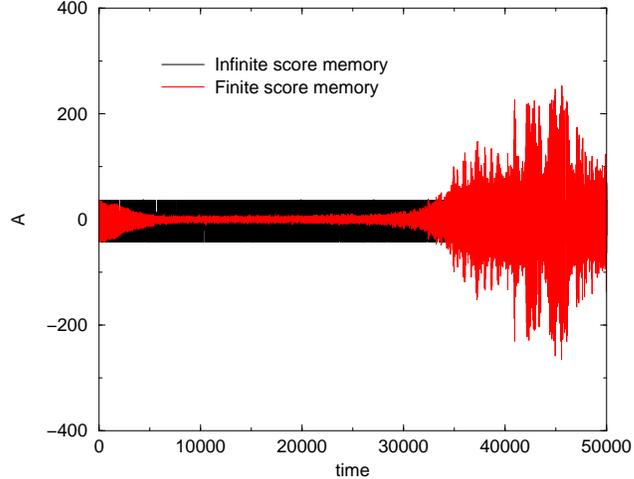}}
\caption{The same game with infinite memory score keeping (red) and
  finite (black). . $PN_s=16000$, $n_s=20$, $n_p=1$, $eps=0.01$}
\label{2runs}
\end{figure}
\begin{figure}
\centerline{\includegraphics[angle=-90,width=0.6\textwidth]{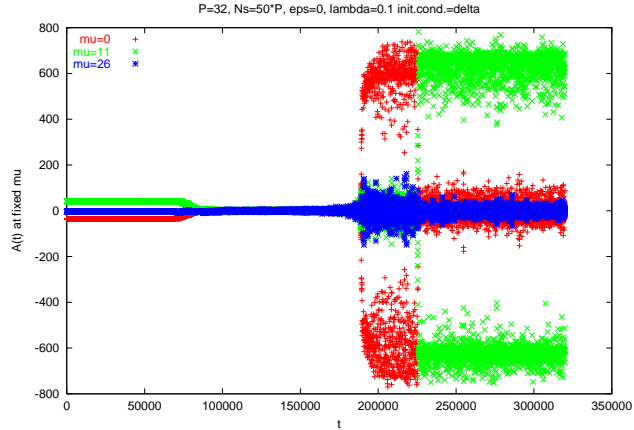}}
\caption{The time series of $A(t)$ at fixed $\mu(t)=0$ (red), $11$
  (green) and $26$ (blue) in a run of the GCMG with
  $P=32,~N_s=1600,~\epsilon=0.01$ and $\lambda=0.1$.}
\label{strange_things}
\end{figure}

The dynamics of the GCMG with finite memory is surprising. At first,
fluctuations are reduced and are Gaussian. Many iterations are needed
before stylized facts emerge again. Remarkably, the typical time for
the emergence of anomalous fluctuations is much larger than
characteristic times, which are of order $P/\lambda$. When one moves
deeper into the crowded market region $N_s/P\gg n_s^c$, one sees even
stronger effects, as shown in Fig. \ref{strange_things}. Still, one
has to wait a quite long time before wild fluctuations set in. But now
large price fluctuations are proportional to the number of speculators
($A(t)\propto N_s$) and they appear associated to a particular
information pattern $\mu$. The sample of Fig. \ref{strange_things}
shows that the value of $\mu$ for which $A(t)\propto N_s$ may even
change in the course of time. What is striking is that a modification
like memory, which naively should make the dynamics smoother by
introducing a finite cutoff in time dependencies, actually provokes
the boost of wild fluctuations. The fact that this effects sets in for
times much longer than the memory decay strongly suggests that its
origin lies in collective fluctuation phenomena. Finally, as for the
plain MG, a finite memory $\lambda>0$ eliminates all dependences on
initial conditions and non-ergodicity effects.

\section{Conclusions}

In short, we have shown that finite score memory suppresses
non-ergodic behavior and dependence on initial conditions in MGs. At
the same time, it brings about new spectacular fluctuation phenomena
in the GCMG. Interestingly, all the analytical tools used so far for
the study of MG fail when $\lambda>0$. On one side, as soon as
$\lambda>0$ the stationary state can no longer be related to the
minimum of a global function, in contrast with the $\lambda=0$ case
\cite{CM03,CMZe99,MC01,MinMaj}. This seems to rule out replica
approaches. On the other, a finite memory destroys the frozen
component of the dynamical variables which allows one to extract
information from the generating functional approach
\cite{CoolenBatch,CoolenOnline}. This makes finite memory MG a quite
challenging problem.


\begin{thebibliography}{99}


\bibitem{Hauert} S. Gy\"orgy and C. Hauert, Phys. Rev. Lett. {\bf 89},
118101 (2002)

\bibitem{CM03} D. Challet and M. Marsili, Phys. Rev. E {\bf 68},
  036132 (2003)

\bibitem{CZ97} D. Challet and Y.-C. Zhang, Physica A {\bf 246}, 407
(1997)

\bibitem{SZ99} F. Slanina and Y.-C. Zhang, Physica A {\bf 272}, 257
(1999)

\bibitem{J99} P. Jefferies, M.L. Hart, P.M. Hui and N.F. Johnson,
Int. J. Th. Appl. Fin. {\bf 3} 3 (2000)

\bibitem{MMM} D. Challet, M. Marsili and Y.-C. Zhang, Physica A {\bf
276}, 284 (2000)

\bibitem{BouchMG2} I. Giardina, J.-Ph. Bouchaud, Eur. Phys. J. B {\bf
31} 421 (2003)

\bibitem{Sornette} J.V. Andersen and D. Sornette, Eur. Phys. J. B {\bf
31}, 141 (2003)

\bibitem{Book} D. Challet, M. Marsili and Y.-C. Zhang, {\it Minority
Games and beyond}, Oxford University Press (Oxford, 2004), in press.

\bibitem{J00} P. Jefferies {\em et al.}, Eur. Phys. J. B {\bf 20}, 493
  (2001)

\bibitem{CMZ00} D. Challet, M. Marsili and Y.-C. Zhang, Physica A {\bf
294}, 514 (2001)

\bibitem{ContBouchaud} R. Cont and J.-Ph. Bouchaud,
Macroecon. Dyn. {\bf 4}, 170 (2000)

\bibitem{MMRZ} M.  Marsili, R. Mulet, F. Ricci-Tersenghi and
  R. Zecchina, Phys. Rev. Lett. {\bf 87}, 208701 (2001)

\bibitem{CMZe99} D. Challet, M. Marsili and R. Zecchina,
Phys. Rev. Lett. {\bf 84}, 1824 (2000)

\bibitem{MC01} M. Marsili, D. Challet, Phys. Rev. E {\bf 64}, 056138
  (2001)

\bibitem{CMM04} D. Challet, M. Marsili, A. De Martino, forthcoming
  (2004)

\bibitem{Muzy} E. Bacry, J.F. Muzy, Commun. Math. Phys. {\bf 236}, 449
(2003)

\bibitem{CZ98} D. Challet and Y.-C. Zhang, Physica A {\bf256}, 514
  (1998)

\bibitem{Jpred} D. Lamper, S. Howison and N.F. Johnson,
Phys. Rev. Lett. {\bf 88}, 017902 (2002)

\bibitem{MinMaj} A. De Martino, I. Giardina and G. Mosetti,
 J. Phys. A: Math. Gen. {\bf 36} 8935

\bibitem{CoolenBatch} J.A.F. Heimel, A.C.C. Coolen, Phys. Rev. E {\bf
63}, 056121 (2001)

\bibitem{CoolenOnline} A.C.C. Coolen and J.A.F. Heimel, J. Phys. A:
Math Gen. {\bf 34}, 10783 (2001)

\end{thebibliography}
\end{document}